# Design and Fabrication of a Low-cost Liquid Optical Waveguide for Augmented Reality

Dechuan Sun[1,2,*], **Gregory Tanyi**[1], **Alan Lee**[1], **Chris French**[2], **Younger Liang**[3], **Christina Lim**[1], **Ranjith R Unnithan**[1,*]

[1]Department of Electrical and Electronic Engineering, The University of Melbourne, Melbourne, Victoria, Australia

[2]Neural Dynamics Laboratory, Department of Medicine, The University of Melbourne, Melbourne, Victoria, Australia

[3]AR Division, KDH Design Taiwan, Neihu Dist, Taipei City 114, Taiwan

**\* Correspondence:**
Dechuan Sun (dechuan.sun@unimelb.edu.au)

Ranjith R Unnithan (r.ranjith@unimelb.edu.au)

## Abstract

The complexities of fabrication techniques and the demand for high precision have posed significant challenges in the mass production of augmented reality (AR) waveguide combiners. Leveraging the capabilities of Polyjet 3D printing techniques, we have developed a cost-effective method for fabricating liquid geometric waveguide combiners for AR applications, using silicone oil as the medium. During the design phase, we optimized the structure of the waveguide combiner to facilitate easier fabrication. Our proposed method simplifies the production process by removing the need for complicated steps like dicing, layer bonding, and polishing, which are usually involved in traditional manufacturing techniques. We conducted optical simulations and developed a prototype using our patented fabrication method, which successfully demonstrated the integration of virtual images with the real-world environment, thereby confirming its feasibility and potential for cost-effective mass production.

# Introduction

Advances in compact mobile devices have led to a rising interest in immersive human-digital interactions that exceed the capabilities of conventional flat-screen displays [1]. As a result, augmented reality (AR) glasses or headsets are emerging as the next generation of interactive devices. These innovative tools enrich the user experience by seamlessly superimposing vivid visual information onto the real world. To date, AR technology has demonstrated significant potential for applications in various domains, such as navigation, engineering, military, gaming, and education [1,2].

AR glasses or headsets incorporate a variety of essential components to provide immersive experiences for users. Among these, the optical see-through near-eye display (NED) stands out as one of the core elements. A NED typically consists of two essential components: a light engine that generates and displays virtual content, and an optical combiner that merges the projected imagery with the user's physical environment. Designing an effective light engine for AR systems requires consideration of key factors such as image brightness, miniaturization, power consumption, heat dissipation, and resolution. To meet these requirements, various micro light engine technologies have been developed and patented, including micro-LED, liquid crystal on silicon (LCoS), micro-organic LED, and digital light processing [3]. In contrast, optical waveguide design and fabrication are intricate and complex processes, posing substantial challenges for engineers and researchers. Lumus, a pioneer in waveguide technology, introduced the first commercially successful geometric waveguide solution in the early 2000s. Their innovative design features a triangular prism for coupling light into the waveguide and integrates an array of transflective mirrors to redirect light toward the user's eyes. Light transmits through the waveguide via total internal reflection, ensuring efficient transmission. The use of total internal reflection within the waveguide ensures efficient light transmission [4,5,6]. Inspired by Lumus's approach, industry leaders such as Vuzix, Microsoft, , Waveoptics and Magic Leap have adopted another design strategy involving the use of diffractive optical elements like gratings [7]. The gratings facilitate efficient coupling of light into and out of the system by diffracting and guiding it along a controlled path within the waveguide [8, 9]. Currently, surface relief gratings have emerged as the most mature technique for fabricating AR waveguides. Another technique known as volume holographic gratings offers more efficient light diffraction, but AR glasses utilizing this approach are still limited in the market [10]. Notable companies at the forefront of this technology include Digilens and Sony. Apart from the previously mentioned techniques, various other methods have been reported. These include the use of metasurfaces, reflective polarization, and photopolymer [11,12,13,14]. While these technologies hold promise, they have not yet been commercialized because of difficulties encountered in the mass production stage.

Both geometric and diffractive waveguides come with their unique sets of benefits and drawbacks [1,2,15]. The fabrication of geometric waveguides is a complex process involving multiple detailed steps. It typically begins with depositing reflective materials onto glass substrates, followed by stacking these coated substrates with adhesive to create a layered structure. The stack is then sliced to the desired thickness, polished for a smooth surface, and cut into the required shapes. Although the fabrication process is complex, this structure offers several advantages, including a large field of view, high optical efficiency, and the absence of color dispersion [1,2]. On the other hand, diffractive waveguides rely on periodic optical structures that can be fabricated using two main techniques. The first method involves fabricating a topographical pattern of peaks and valleys on a thin glass substrate to modulate light propagation. The second approach leverages the interference of laser beams, generating periodic bright and dark fringes for efficient light guidance within the waveguide [1]. In contrast to the complex fabrication process of geometric waveguides, which involves layer bonding, slicing, and polishing steps, the production of diffractive waveguides is relatively simpler. However, the primary limitation of diffractive waveguides is their susceptibility to color dispersion, commonly known as the "rainbow effect" [1,2,16]. This phenomenon arises from the sensitivity of periodic optical structures to the angle and wavelength of the input light, making it challenging to achieve consistent optical performance across the visible spectrum. Although both geometric and diffractive waveguides hold significant potential for AR applications, their widespread commercialization is hindered by limitations in mass production capabilities.

Recent advancements in PolyJet 3D printing technology have significantly transformed the field of additive manufacturing [17,18]. Unlike conventional UV resin 3D printing, PolyJet printing uses water-soluble support materials, which result in minimal surface blemishes on the finished part. This capability allows for the printing of overhangs without compromising the overall quality of the printed surface. In this paper, we first designed and optimized a geometric waveguide with a focus on fabrication feasibility.

Subsequently, utilizing our patented manufacturing processes, we successfully fabricated a cost-effective liquid waveguide using silicone oil and a 3D-printed waveguide frame that integrated three dielectric reflectors for AR applications [19]. Finally, we developed a custom-designed automated waveguide assembly system for future use. Our proposed method has the potential to streamline the waveguide manufacturing process by eliminating the need for complex machining processes. This simplification can lead to shorter lead times and increased cost-efficiency, making mass production more feasible. As far as we know, the use of liquid materials as an AR waveguide has not been previously reported, and it offers numerous advantages. The high optical clarity of the liquid across a broad wavelength band minimizes scattering losses and improves overall performance. Unlike solid materials, liquids can conform to the shape of the waveguide without applying additional stress or causing distortion. Using liquid as the medium eliminates reflector bending issues commonly encountered during the layer bonding process in conventional geometric waveguide fabrication. Additionally, liquid waveguides offer the advantage of self-healing minor surface defects, which improves the durability and reliability of the optical system. These characteristics position liquid-based waveguides as promising candidates for next-generation AR waveguides.

## Waveguide design and simulation

The waveguide comprises a 3D-printed frame with integrated dielectric reflectors, sealed with thin cover glass pieces, and filled with silicone oil. Using the finite element method and the ray optics module in COMSOL Multiphysics, we designed and optimized the geometric waveguide structure to ensure good optical performance for AR applications while also considering ease of fabrication.

The waveguide we designed has dimensions of 38mm in length, 28.5mm in width, and 3.2mm in thickness. If the thickness of the waveguide is too thin, it may lead to reflector alignment issues during the later assembly stage. To couple the projected image into the waveguide, our design features a triangular prism structure with a 50-degree slant angle relative to the waveguide's base. This slant angle ensures efficient light transmission in the waveguide through total internal reflection, maintaining minimal light loss. Three dielectric reflectors are positioned at a 25-degree angle relative to the waveguide's base to direct light toward the user's eyes. These reflectors are arranged in parallel with a distance of 2.024mm between each pair. Based on measurements, a 90% transmittance ratio is used in the simulation to account for propagation loss, with wall boundary conditions applied. Three thin cover glass pieces, each with a thickness of 0.1 mm and a refractive index of 1.46, are placed on the top and bottom surfaces of the waveguide, as well as on the surface of the triangular prism. The waveguide is filled with silicon oil, and we use a refractive index of 1.41 for the optical simulation. The design utilizes a standard LCoS projector that projects a virtual image through a series of lenses into the waveguide via the triangular prism. To model the human eye, we represent the crystalline lens as an ideal convex lens (focal length: 25mm) with optimal lens boundary conditions applied. Additionally, an image plane is positioned 25mm behind this lens to simulate the retina.

In order to assess the performance of our waveguide design, we selected a Super Mario image as the projected input and examined its reconstruction on the simulated retina. The image was converted to the bitmap file format and imported into the projector module. Thus, the spatial and energy distributions of the rays emitted from the object were aligned with the bitmap data. Using 125000 tracing rays, the designed waveguide successfully reconstructed a perfectly stitched image. Figure 1a-b illustrates the simulated ray tracing results, providing a graphical representation of the rays' transmission within the waveguide combiner and the convergence of light on the retina.

## Fabrication protocol

The entire design of the waveguide is shown in Figure 2a. Utilizing the Stratasys J826 PolyJet 3D printer, we first 3D printed a waveguide frame that replicated the design of the simulated waveguide. To enhance printing quality, a fluorinated ethylene propylene film or a piece of flat smooth glass was adhered to the printing bed. In order to accommodate the insertion of dielectric reflectors, we incorporated 200μm-wide slots into the side walls of the frame. This slot width was determined through a series of tests with 5μm increments from 165μm to 200μm. These tests revealed that 180 μm provided the optimal balance between ease of insertion and alignment accuracy. The thickness of the reflectors is 175 μm. To facilitate the subsequent silicone oil injection and prevent bubble formation, 0.5mm venting holes were incorporated into the side wall of the frame. We selected the transparent resin RGD810 for its rigid transparency and resistance to deformation, which features a minimum flexural strength of 75 MPa. To facilitate the easy removal of support structures, we utilized a water-soluble resin SUP707. The frame structure was printed in high-quality mode with a z-axis resolution of 14 μm. The printing temperature was maintained at 22°C, and the humidity was around 35%. The sample was printed with a glossy surface finish and only one UV lamp was enabled during the printing. The entire printing process took approximately 3 hours to complete. After printing, a solution containing 2% sodium metasilicate and sodium hydroxide was utilized to accelerate the dissolution of the support material without damaging the main structure. The sample was immersed in the solution within a vibration cleaning station for a duration of 1 hour and then washed using a high-pressure water gun. Residual support material in the slots and holes was carefully removed using a razor blade and a needle. The sample was then rinsed thoroughly using deionized water. Finally, the sample underwent 5 minutes of photobleaching with UV floodlight to enhance transparency and reduce yellowing.

Prior to inserting the reflectors, silicone oil was applied to the slots to reduce friction and ensure smooth insertion. Using precision tweezers, three dielectric filters were delicately inserted into the designated slots of the frame, with the process closely monitored under a microscope. A common method to assess the alignment of reflectors involves utilizing a laser and measuring the distance between the reflected spots from each pair of reflectors. However, a more convenient approach is to use ceiling light to visually inspect the reflected image (Figure 2b). This simplified technique allows for quick and easy alignment verification without the need for additional equipment or complex measurements. Next, the cover glass was prepared by cutting it to the appropriate size using a dicing saw. The glass was then subjected to plasma cleaning to remove any debris or contaminants before being placed on a flat glass bed within a nitrogen-filled chamber. Following this, a thin layer of UV epoxy was applied to the bottom surface of the waveguide frame, which was then positioned onto the cover glass for bonding. UV light was directed through the bottom of the glass bed to cure the UV epoxy with appropriate pressure applied on the top of the frame. To prevent excess epoxy from adhering to the glass bed, a thin film of polytetrafluoroethylene was deposited on the glass bed, providing a non-stick surface for easier cleanup and maintenance. This process was repeated to bond the cover glass to the top side of the frame and the surface of the triangular prism. After completing the bonding process, a 27-gauge needle and a 5 ml syringe connected to a syringe pump were used to inject silicone oil into the frame chamber. The silicone oil, with a viscosity of 20 cSt, was injected at an optimal flow rate of 0.2 ml/min, determined through a series of tests to achieve the most stable results. It is crucial to avoid injecting the oil too rapidly, as this can cause the formation of bubbles, potentially compromising the waveguide's overall performance. The silicone oil has a density of 0.95 g/mL, whereas the UV resin has a higher density of 1.1 g/mL. To seal the chamber after filling it with silicone oil, UV resin was swiftly applied to the hole and cured with UV light to prevent the resin from sinking due to its higher density.

## Results

During the fabrication process, dielectric bandpass filters were used as reflectors. We evaluated the performance of the reflectors by measuring the transmittance ratio at two angles using a Cytoviva hyperspectral microscope across the visible light spectrum. The transmittance ratio was determined by averaging data obtained from 20 distinct test points on the sample. The selected angles were 0 degrees, representing the angle at which light is projected perpendicularly to the reflector, and 25 degrees, corresponding to the input light angle from the projector. The transmittance ratio remained relatively stable at both degrees, with an approximate 3% decrease observed across the 400nm to 700nm wavelength range at 0 degrees. Additionally, the transmittance ratio at 0 degrees was approximately 2% higher on average compared to the ratio at 25 degrees. (Figure 3a). These findings indicate that integrating reflectors into the design does not significantly compromise the overall color fidelity of the system. Subsequently, we used an atomic force microscope (Bruker Dimension Icon) to measure the surface roughness of the reflectors. The root mean squared roughness ($R_q$) was determined to be 1.4 nm, and the mean roughness ($R_a$) was measured at 1.09 nm. The cover glass piece we used has a thickness of 0.1 mm and a refractive index of 1.46. It maintains a consistent transmittance ratio of approximately 93% at an angle of 0 degrees throughout the 400nm to 700nm wavelength range (Figure 3b). Surface roughness analysis revealed a root mean squared roughness ($R_q$) of 1.3 nm and a mean roughness ($R_a$) of 1.13 nm, ensuring minimal light scattering and optimal optical performance.

To evaluate the performance of the proposed design, we constructed a waveguide prototype utilizing the proposed fabrication methods described above. The waveguide is remarkably lightweight, weighing only 5.4 grams, with an overall prototype cost of approximately $18. However, mass production can lead to a significant reduction in cost. The transmittance ratio curve of the waveguide consistently remains above 77% within the visible light wavelength range (Figure 3c), and it demonstrates outstanding transparency in both near and long-range viewing conditions, as illustrated in Figure 4a-b. In the experimental setup, an LCoS micro-projector is precisely aligned and positioned at the triangular coupling surface of the waveguide for optimal functionality. The entire experimental setup is shown in Figure 4c. Our waveguide prototype effectively guides and projects images from the micro-projector to the user's eye. Figure 4d shows an example of the digital image loaded into the projector and the corresponding virtual image that overlaps with the laboratory environment as observed through the waveguide. Additionally, a supporting video demonstrating the successful image reconstruction by the waveguide is included in the supplementary material.

Next, we studied the optical properties of the fabricated waveguide. For evaluating the field of view (FOV) and the eye box of the AR system, an eye relief distance of 10 mm was chosen as the standard. To measure the projection length of the reconstructed image, a target board was placed 25 cm away from the eye pupil plane. The lengths of the projected image were measured in both the horizontal (86 mm) and vertical (55 mm) directions on the target paper. These measurements correspond to a horizontal field of view (FOV) of approximately 19.52 degrees and a vertical FOV of approximately 12.56 degrees. Since the primary objective of this study was to assess the feasibility of the design and the fabrication method, our design includes only three reflectors for simplicity, resulting in a limited eye box of approximately 1 mm. However, incorporating additional reflectors can further expand the eye box and FOV. The properties of the waveguide are summarized in Table 1.

## Discussions and conclusions

In this study, we employed PolyJet 3D printing technology to develop a cost-effective liquid optical waveguide for augmented reality applications, using silicone oil as the waveguide medium. Inspired by the geometric AR waveguide designs, we optimized the waveguide structure to streamline the fabrication process, achieving a more efficient and feasible production method. This optimization eliminated complex and time-consuming procedures such as dicing, layer bonding, and polishing, which are typically required in conventional manufacturing methods. Our fabrication method demonstrates significant potential for adaptation in large-scale production applications.

For simplicity, we choose glass as the material to seal the waveguide chamber, which exhibits a transmittance ratio of approximately 93%. It is worth noting that this transmittance ratio can be further improved through the deposition of anti-reflective coatings like $MgF_2$ or $TiO_2/SiO_2$ [20,21]. Despite these benefits, glass remains highly susceptible to scratching and damage. In contrast, sapphire is a compelling alternative due to its exceptional hardness, which makes it highly resistant to scratching. Additionally, its superior strength and durability make it an excellent choice for this application.

While general UV 3D printers are capable of producing intricate small features, they face limitations with designs that incorporate numerous overhanging structures. Adding support material can assist with printing but can significantly impair reflector alignment. Without support material, these printers are unable to accurately reproduce the intended design (Figure 5a). A potential solution to this challenge could involve developing a novel printing resin that combines exceptional mechanical strength with reduced adhesion properties. On the other hand, although PolyJet printers are more expensive, they offer several advantages, such as an exceptionally large printing plate (255 mm x 252 mm) that allows for the simultaneous printing of multiple waveguide frames. Additionally, since the waveguide does not involve intricate features, the printing time can be further reduced by adjusting the printing resolution with proper optimization. Throughout the waveguide design and fabrication processes, we have ensured that they meet mass production requirements. To further streamline the fabrication process, we collaborated with FiconTEC and designed an automatic waveguide assembly system based on the proposed fabrication procedures, as illustrated in Figure 5b. We are currently evaluating this system's performance and capabilities, aiming to utilize it for future automated waveguide assembly tasks.

In conclusion, we have developed a prototype of the AR waveguide using our patented fabrication methods. The key advantage of this prototype lies in its ability to bypass complex steps such as dicing, layer bonding, and polishing, which are usually necessary in conventional manufacturing processes. This innovative waveguide effectively merges virtual images with real-world surroundings, demonstrating its feasibility and potential for cost-effective mass production.

## Acknowledgments

The project was funded and supported by JARVISH. Additionally, this work was conducted in part at the Melbourne Centre for Nanofabrication (MCN) in the Victorian Node of the Australian National Fabrication Facility (ANFF).

## Author Contributions

Conceptualization, D.S.; simulation, G.T.; methodology, D.S. and A.L.; original draft, D.S.; supervision, Y.L., C.L., and R.R.U.; experimental resources, C.F., Y.L., C.L., and R.R.U.

## Conflict of interest

All authors disclose no conflict of interest for this work.

## Data availability statement

The data that support the findings of this study are available from the corresponding author upon reasonable request.

# Figures

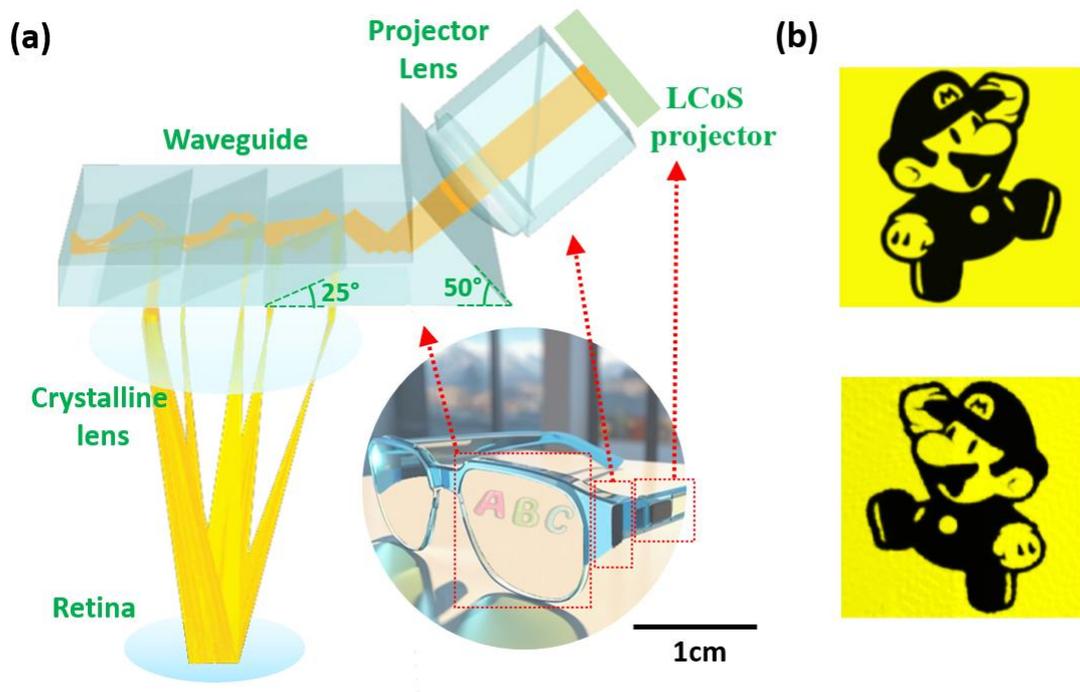

**Figure 1**. **An optical simulation of the proposed waveguide, illustrated by a ray tracing diagram, demonstrates the light path, image formation, and pupil expansion. (a)** A virtual image from the projector enters the waveguide through the projector lens and a triangular prism. The light propagates through the waveguide until it encounters dielectric reflectors, which allow it to exit. Finally, the crystalline lens focuses the light, forming a clear image on the retina. **(b)** The top subfigure displays an example of the projected image, while the bottom subfigure shows the reconstructed image.

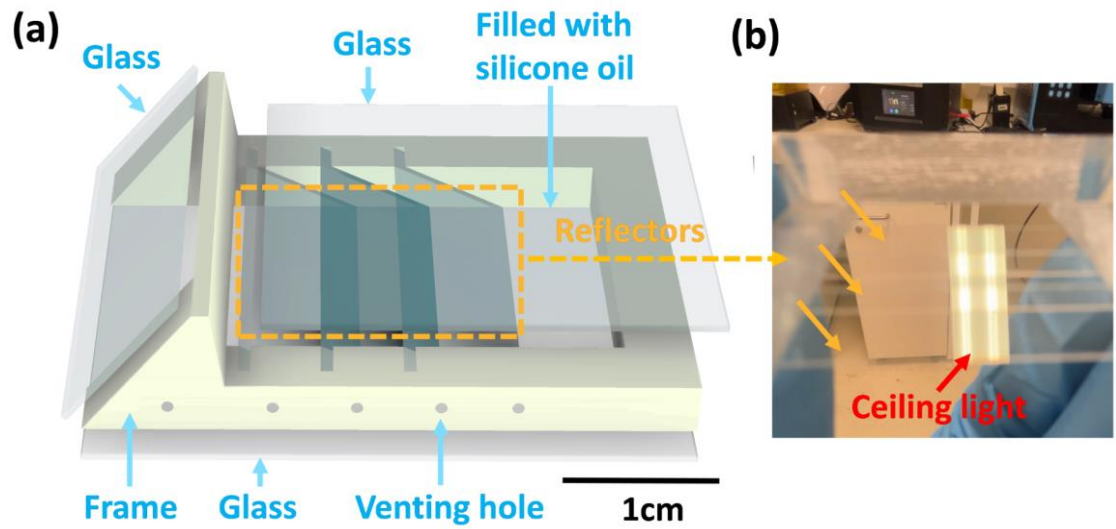

**Figure 2. The fabrication of the proposed waveguide.** **(a)** A structural diagram of the proposed waveguide. **(b)** A 3D-printed frame integrates three reflectors. The alignment of the reflectors is checked using the ceiling light.

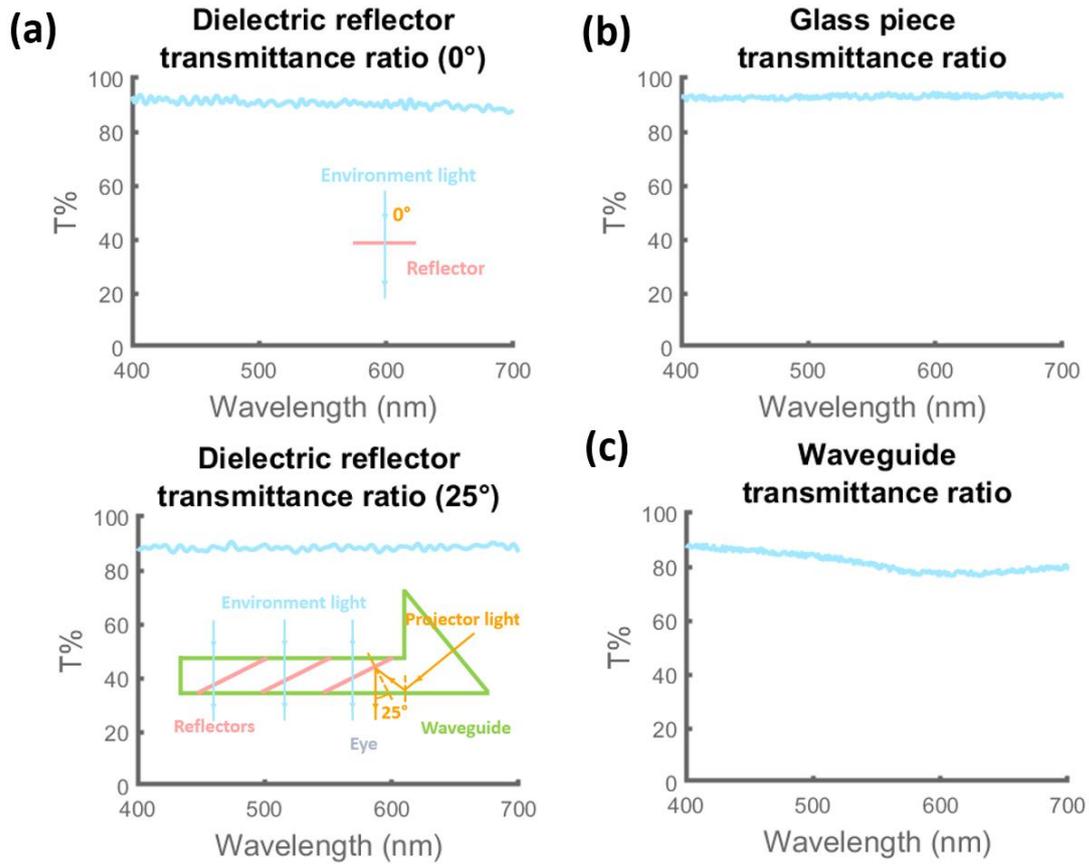

**Figure 3. Measurement of the transmittance ratio. (a)** The transmittance ratio of the dielectric filter was measured at angles of 0 and 25 degrees. A 25-degree angle was chosen to simulate the incident light angle of the projector. **(b-c)** The transmittance ratio of the glass piece and the assembled waveguide.

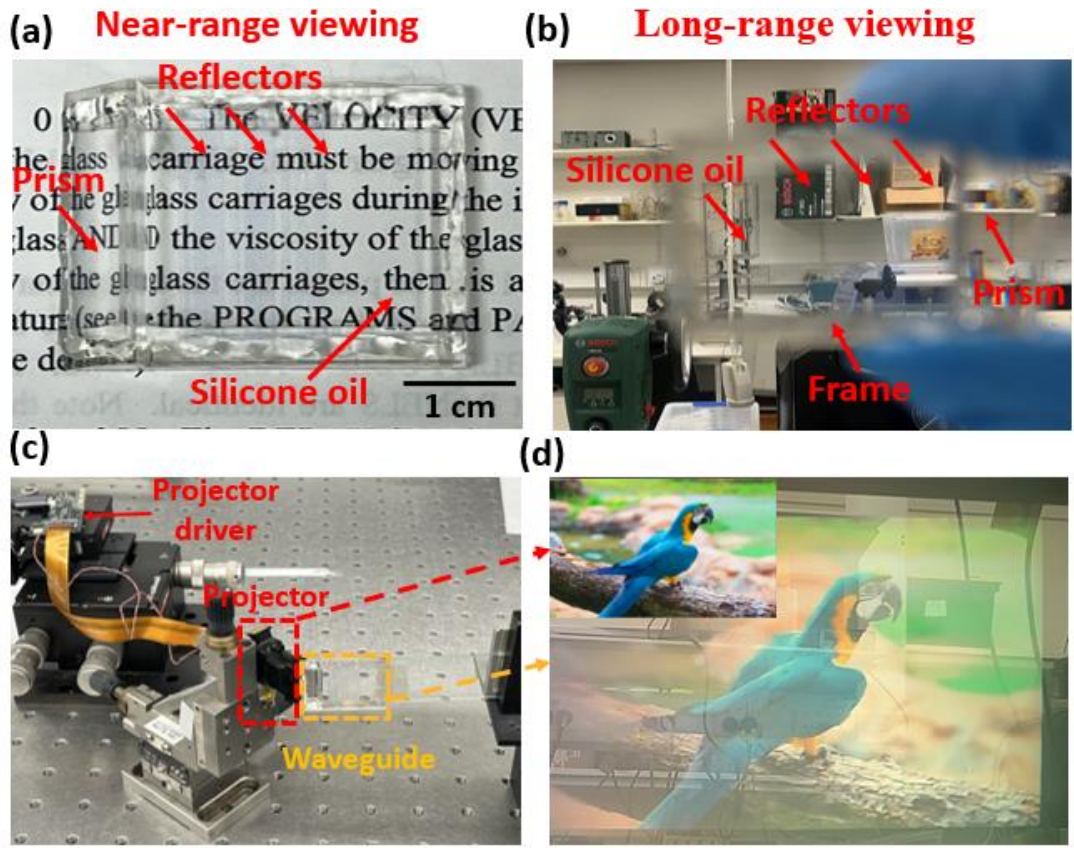

**Figure 4. Optical performance of the fabricated waveguide. (a-b)** The assembled waveguide exhibits excellent transparency in both near and long-range viewing conditions. Extra UV epoxy was applied for better sealing on edges. **(c)** Experimental setup. The waveguide is adhered to a piece of holding glass, and the projector is precisely aligned with the triangular prism. **(d)** A demonstration of the projected image alongside the reconstructed virtual image that overlaps with the laboratory environment.

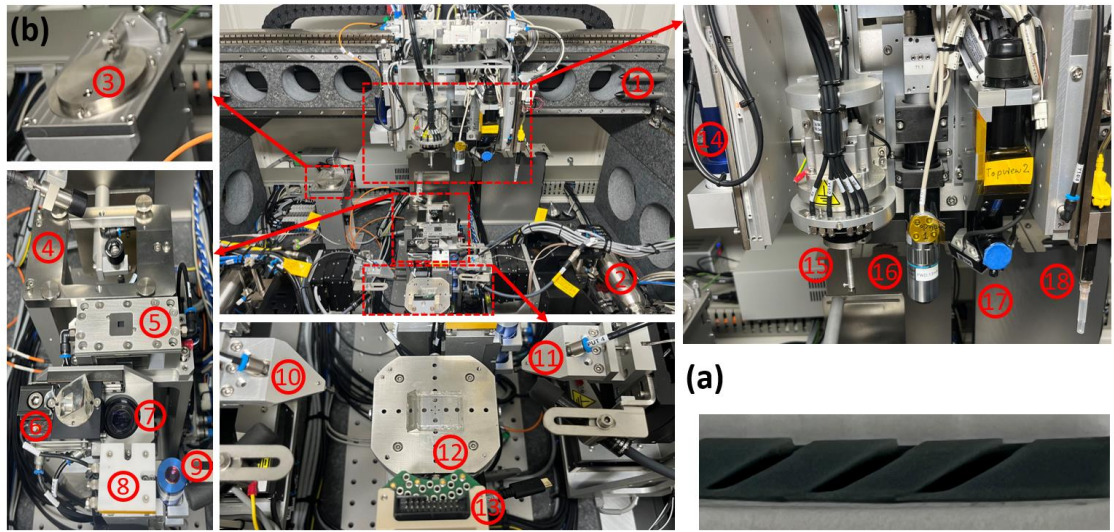

**Figure 5. A sample printed using an LCD UV printer and a custom-designed automated waveguide assembly system.** **(a)** A general LCD UV printer cannot precisely print overhanging structures without using support materials (designed slot width:150um, 200um, 250um). **(b)** Our custom-designed automatic waveguide assembly system. 1:gantry system; 2:damping system; 3:UV glue preparation platform; 4:laser for precise 3D-printed frame cleaning; 5:heat plate to help relieve accumulated stress during UV glue curing; 6:sideview lens; 7:bottom view camera; 8:glass piece preparation platform; 9:distance measurement sensor; 10-11: reflector pickers with pressure sensors integrated; 12:sample preparation platform; 13:light engine interface; 14:UV light; 15:wavguide picker; 16-17:top view camera with different magnification; 18:silicone oil syringe.

| | |
|---|---|
| Micro-projector | LCoS |
| Fabrication cost | $18 |
| Dimensions | 38mm x 28.5mm x 3.2mm |
| Weight | 5.4g |
| Waveguide structure | Geometric (with 3 reflectors) |
| Refractive index | 1.41 |
| Transmittance ratio | Above 77% within the visible light wavelength range |
| Eye relief distance | 10 mm |
| FOV | Horizontal: 19.52°<br>Vertical: 12.56° |
| Eye box | 1 mm |

**Table 1. Properties of the proposed AR waveguide.**